\documentclass[cameraready]{Interspeech}

\title{Time-Unconditional Generative Speech Enhancement via Autonomous Rectified Flow}

\author[affiliation={1}, orcid=0009-0006-2552-1234]{Wen}{Zhang}
\author[affiliation={1}, orcid=0000-0002-4063-8952, correspondingauthor]{Wenbin}{Jiang}
\author[affiliation={1}, orcid=0009-0006-1175-0749]{Yang}{Zhang}
\author[affiliation={2}, orcid=0000-0002-7977-9728]{Xiaofei}{Zhou}

\address{
    $^1$School of Communication Engineering, Hangzhou Dianzi University, Hangzhou 310018, China
    $^2$School of Automation, Hangzhou Dianzi University, Hangzhou 310018, China 
}

\email{zhangwen@hdu.edu.cn, wbjiang@hdu.edu.cn}

\usepackage{comment}
\usepackage{booktabs,multirow}
\begin{document}

\maketitle

\begin{abstract}
Most generative speech enhancement methods rely on explicit time-step embeddings for temporal conditioning. In this paper, we propose the Autonomous Rectified Flow framework, which challenges the necessity of such conditioning. Using a linear interpolation path, we show that the target vector field is inherently time-invariant. We further introduce a time-unconditional network that eliminates explicit time-step information and infers the denoising direction solely from the spatial relationship between the current state and the noisy observation. Predicting this target vector field is equivalent to modeling the noise distribution. By avoiding overfitting to temporal trajectories, the proposed autonomous design significantly improves generation quality, robustness, and inference efficiency.
\end{abstract}

\section{Introduction}

Speech enhancement (SE) is a fundamental task in speech processing that aims to recover clean speech from degraded observations. Classical methods, such as spectral subtraction and Wiener filtering, are based on statistical signal processing techniques~\cite{loizou2007speech}. The advent of deep learning has revolutionized speech enhancement, with deep neural network (DNN)-based methods becoming the dominant approach and achieving substantial improvements in perceptual quality and intelligibility~\cite{hendriks2013dft,jiang2023,jiang_unse_taslp,jiang2026selfse}. However, as discriminative models, they often struggle to generalize to unseen acoustic conditions and adverse environments.

Recently, generative paradigms, particularly diffusion probabilistic models and score-based generative models, have established new benchmarks in SE by effectively modeling the complex distribution of speech signals~\cite{welker22_interspeech,10149431,ho2020denoising,Lemercier_2023}. However, their practical deployment is hindered by two key challenges: the high computational cost of iterative sampling and the prior mismatch problem. The latter stems from the fact that standard diffusion processes transform clean speech into an uninformative Gaussian prior, whereas the noisy observation $y$ in SE provides a more structured and task-relevant prior.

To mitigate this prior mismatch, recent research has shifted toward boundary-anchored stochastic processes, such as Brownian Bridge with Exponential Diffusion (BBED)~\cite{lay23_interspeech}, which constrains the generative trajectory to lie between clean speech and the noisy mixture. Similarly, Flow Matching (FM)~\cite{lipman2023flow,liu2023flow} and FlowSE~\cite{10888274,zhang2026hyflowse} introduce deterministic ordinary differential equation (ODE)~\cite{chen2018neural} trajectories based on optimal transport. By employing straight-line paths, these frameworks reduce the Number of Function Evaluations (NFEs) required for reverse-time integration while maintaining high-fidelity speech reconstruction. However, despite these advances, existing models still rely on explicit time-step embeddings to modulate the vector field.

While explicit temporal conditioning is essential in standard diffusion models for tracking the evolving noise scale, we argue that it is unnecessary—and potentially harmful—in boundary-anchored SE~\cite{pmlr-v267-sun25g}. When the generative path is linear~\cite{liu2023flow}, the target velocity vector remains constant throughout the entire process. Consequently, time-conditioned models must learn trajectory-specific noise scales, making them prone to trajectory overfitting. During inference, even small numerical deviations can cause the input state to depart from the expected distribution at a given time step, reducing the model's ability to recover. In contrast, modeling the enhancement process as an autonomous dynamical system enables the vector field to be fully independent of explicit temporal variables.

\begin{figure}[t]
  \centering
  \includegraphics[width=0.75\linewidth]{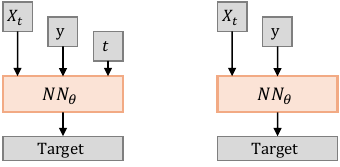}
  \caption{This work investigates the removal of explicit time step $t$ in generative speech enhancement models, where $X_t$ denotes the linear interpolation of clean speech and noisy speech, $y$ represents the noisy speech input, and $t$ is the time step.}
  \label{fig:model}
  \vspace{-12pt}
\end{figure}

In this paper, we propose the Autonomous Rectified Flow (ARF) framework for time-unconditional SE, bridging the gap between flow matching and direct noise prediction~\cite{10887784}. The main contributions of this paper are summarized as follows:
\begin{itemize}
\item We reveal the time-invariance of the target vector field in linear-path generative SE and prove that explicit time-step embeddings are mathematically redundant for this task.
\item We propose ARF, a novel time-unconditional SE framework built upon an autonomous ordinary differential equation (ODE) system, which completely removes explicit temporal conditioning(Figure~\ref{fig:model}) and learns a static denoising direction equivalent to the underlying noise distribution.
\item ARF enables highly efficient inference. It achieves a competitive PESQ score of 3.11 with $NFE=5$ and reduces the RTF to as low as 0.02 with $NFE=1$, significantly outperforming existing single-step generative SE methods.
\end{itemize}

\section{Related Work}
\subsection{BBED: Brownian Bridge with Exponential Diffusion}
To mitigate the prior mismatch inherent in score-based generative models for SE, we adopt a forward process based on a Brownian bridge~\cite{lay23_interspeech}. Unlike conventional stochastic differential equations (SDEs), whose means converge to the noisy mixture only in the limit, a Brownian bridge is characterized by fixed boundary conditions. Using a linear interpolation factor $k(t)=t$ for $t\in[0,1)$, the mean of the SDE is defined as:
\begin{equation}
    \mu(t) = (1 - t)x_0 + ty
\end{equation}
According to the BBED framework, the specific drift coefficient $f(x_t, y)$ required to achieve this mean evolution is derived by solving the corresponding SDE equations:
\begin{equation}
    f(x_t, y) = \frac{y - x_t}{1 - t}
\end{equation}
Combined with an exponential diffusion coefficient $g(t) = \sqrt{c}k^t$, this formulation ensures that the variance $\sigma(t)^2$ vanishes at both $t=0$ and $t=1$, effectively pinning the process to the clean speech $x_0$ and the noisy mixture $y$.

To learn the reverse process for enhancement, the score model $s_{\theta}(x_t, y, t)$ is optimized based on denoising score matching. The training target aims to minimize the following objective function:
\begin{equation}
    \arg \min_{\theta} \mathbb{E}_{t, (x_0, y), z} \left[ \left\lVert \sigma(t) s_{\theta}(x_t, y, t) + z \right\rVert_2^2 \right]
\end{equation}
where $x_t = \mu(t) + \sigma(t)z$ with $z \sim \mathcal{N}_{\mathbb{C}}(0, I)$. This training scheme allows the network to approximate the score function $\nabla_{x_t} \log p_t(x_t | y)$, which is then used to solve the reverse SDE to generate an estimate of the clean speech $x_0$ from $y$. By reducing the prior mismatch through the BBED structure, this training objective leads to more stable and accurate speech recovery than baseline methods.

\subsection{FlowSE: Flow Matching-based Speech Enhancement}
To address the heavy computational complexity problem in traditional diffusion models~\cite{10888274}, this methodology adopts Conditional Flow Matching (CFM) to train Continuous Normalizing Flows (CNFs). FlowSE models the probability path from a noisy observation to the clean speech distribution through a deterministic ODE:
\begin{equation}
    \frac{d\phi_{t}(x_{0}|y)}{dt}=v_{t}(\phi_{t}(x_{0}|y)|y)
\end{equation}
where $\phi_{0}(x_{0}|y)=x_{0}$ represents the initial state sampled from a conditional prior $p(x_0|y)$, and $v_t$ denotes the time-dependent vector field. Inspired by optimal transport, the framework defines a Gaussian conditional probability path with a linear mean and standard deviation:
\begin{equation}
    \mu_{t}(x_{1},y) =tx_1 + (1-t)y, \quad \sigma_{t} = (1-t)\sigma
\end{equation}
This formulation ensures that the mean moves linearly from the noisy speech $y$ at $t=0$ to the clean speech $x_1$ at $t=1$, effectively eliminating the prior mismatch. For this defined Gaussian conditional path, the conditional flow and the target conditional vector field are given in their general closed forms:
\begin{equation}
    \phi_{t}(x_{0}|x_{1},y)=\frac{\sigma_{t}}{\sigma_{0}}(x_{0}-\mu_{0}(x_{1},y))+\mu_{t}(x_{1},y)
\end{equation}
\begin{equation}
    v_{t}(x_{t}|x_{1},y)=\frac{\sigma_{t}^{\prime}}{\sigma_{t}}(x_{t}-\mu_{t}(x_{1},y))+\mu_{t}^{\prime}(x_{1},y)
\end{equation}
where $\sigma_{t}^{\prime}=\frac{d}{dt}\sigma_{t}$ and $\mu_{t}^{\prime}=\frac{d}{dt}\mu_{t}$. During training, the intermediate state is explicitly sampled via $x_t = \phi_t(x_0 | x_1, y)$. The velocity vector field network $v_{\theta}(x_{t},y,t)$ is optimized by minimizing the CFM loss function:
\begin{equation}
    \mathcal{L}_{\text{CFM}}(\theta) := \mathbb{E}_{t,x_0,x_1,y} \bigl\lVert v_{\theta}(x_t,y,t) - v_{t}(x_t | x_1, y) \rVert^2
\end{equation}
In the inference stage, the clean speech estimate is reconstructed by solving an ODE from $t=0$ up to $t=1$ across $N$ discrete time points $0 = t_0 < t_1 < \dots < t_{N-1} = 1-t_\delta < t_N = 1$, where a small positive threshold $t_\delta$ avoids the endpoint singularity. Using a numerical solver such as the Euler method, the state is iteratively updated along the learned vector field:
\begin{equation}
    x_{t_i} = x_{t_{i-1}} + v_{\theta}(x_{t_{i-1}}, y, t_{i-1}) \Delta t(i)
\end{equation}
where the step size schedule is defined as $\Delta t(i) = \frac{1 - t_\delta}{N-1}$ for $i \in \{1, \dots, N-1\}$ and $\Delta t(N) = t_\delta$. By following these straight-line deterministic trajectories, FlowSE achieves high-fidelity speech recovery with as few as 5 function evaluations ($NFE=5$), significantly enhancing inference efficiency compared to stochastic diffusion processes.

\begin{figure*}[t]
  \centering
  \includegraphics[width=0.95\linewidth, trim=0pt 5pt 0pt 5pt, clip]{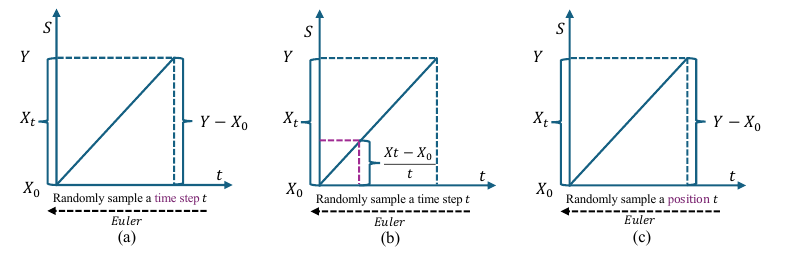}
  \caption{(a) denotes Rectified Flow, (b) denotes Conditional Flow Matching, and (c) denotes Autonomous Rectified Flow. We observe that the training objective of Rectified Flow is fixed and independent of the time step $t$, thus we propose (c), where $t$ represents the current position instead of the time step embedding of $t$. Where $X_t$ denotes the linear interpolation of clean speech and noisy speech, $Y$ represents the noisy speech input, $X_0$ represents the clean speech input.}
  \label{fig:comparison}
  \vspace{-10pt}
\end{figure*}

\section{Proposed Approach}
We illustrate the conceptual differences between our method and prior approaches in Figure~\ref{fig:comparison}. As shown in Figure~\ref{fig:comparison}(a)–(b), traditional Rectified Flow and Conditional Flow Matching methods rely on explicit time-step embeddings or noise schedules to guide the network in predicting the denoising direction. In contrast, high-dimensional speech signals inherently contain sufficient information for inferring the noise level from a noisy observation~\cite{pmlr-v267-sun25g}. Leveraging this property, we propose the Autonomous Rectified Flow framework (Figure~\ref{fig:comparison}(c)), which removes explicit temporal conditioning and predicts the denoising direction solely from the current state and the noisy mixture.

\subsection{ARF: Autonomous Rectified Flow}
The ARF framework models the generative process by leveraging the time-invariance of a denoising target vector field. For a clean speech $x_0$ and a noisy observation $y$, we define a straight path connecting $x_0$ and $y$:

\begin{equation}
    x_t = (1 - t)x_0 + t(y + \sigma z), \quad z \sim \mathcal{N}(0, I)
\end{equation}
where $z$ is Gaussian noise used for stochastic regularization. The target update at each intermediate state is then:
\begin{equation}
    u = \frac{d}{dt}x_t = (y + \sigma z) - x_0
\end{equation}
Because the noisy mixture $y$ is mathematically defined as the clean speech $x_0$ plus additive acoustic noise $n$ (i.e., $y = x_0 + n$), we can rewrite the target vector field as:
\begin{equation}
    u = (x_0 + n + \sigma z) - x_0 = n + \sigma z
    \label{eq:u}
\end{equation}
This shows that the target is determined solely by the noise realization, independent of any temporal scaling. We define the neural network to directly estimate the target vector field:
\begin{equation}
    \mathcal{L}_{\text{ARF}}(\theta) = \mathbb{E}_{t, x_0, y, z} \left[ \left\lVert v_\theta(x_t, y) - u \right\rVert^2 \right]
\end{equation}
During training, the network learns to map the current noisy state to the corresponding noise correction, without relying on any explicit time step embedding. This avoids overfitting to temporal patterns and encourages smoother, more generalizable denoising fields.

\subsection{Autonomous ODE Solver}
Since the learned vector field $v_\theta(x_t, y)$ does not depend on an external time variable, the generation process is governed by an Autonomous Ordinary Differential Equation (ODE). To recover clean speech, we initialize the system at $t=1$ using the noisy prior and integrate backward to $t=0$:
\begin{equation}
    \frac{d\phi_{t}}{dt} = v_{\theta}(\phi_{t}, y), \quad \phi_1 = y + \sigma z
\end{equation}
It should be noted that although $v_\theta$ does not take $t$ as an explicit input, the predicted velocity evolves dynamically along the trajectory because the state $\phi_t$ changes at each integration step. To numerically solve this autonomous ODE, we adopt a multi-step Euler solver with a uniform sampling schedule. The integration points $1 = t_0 > t_1 > \cdots > t_N = 0$ are uniformly distributed as:
\begin{equation}
    t_i = 1 - \frac{i}{N}, \quad i \in \{0, 1, \dots, N\}
\end{equation}
The state is iteratively updated backward with a constant step size $\Delta t = \frac{1}{N}$:
\begin{equation}
    x_{t_{i+1}} = x_{t_i} - v_\theta(x_{t_i}, y) \Delta t
\end{equation}
This approach ensures a deterministic number of network evaluations while fully exploiting the information contained in the noisy observation, without requiring explicit knowledge of the noise schedule or time steps.

\section{Experiments}
\subsection{Datasets}
We conduct experiments on the VoiceBank+DEMAND~\cite{botinhao2016investigating} dataset, which consists of clean speech from VCTK~\cite{veaux2013voice} mixed with noise from DEMAND~\cite{thiemann2013diverse}. In addition, to evaluate the generalization capability of the proposed model, we additionally employ the public synthetic test set from the INTERSPEECH 2020 Deep Noise Suppression (DNS) Challenge~\cite{reddy20_interspeech} for a comprehensive evaluation.

\subsection{Implementation Details}
Our model is based on the Noise Conditional Score Network (NCSN++)~\cite{lemercier2023storm} architecture. Specifically, we freeze the time step input $t$ and the noise scheduling module to construct our time-unconditional autonomous network. We benchmark our framework against two generative SE baselines, FlowSE and BBED; all three models adopt a consistent 65.6M parameter architecture. For ablation studies, the models are trained under the same framework with a unified parameter size of 27.8M for consistent comparison. Results of FlowSE are obtained using publicly available pre-trained models for inference.

All models are trained with the Adam optimizer with a learning rate of $1 \times 10^{-4}$ and a batch size of 4 for 100 epochs. The parameter $\sigma$ is set to 0.5, and Exponential Moving Average (EMA) is applied to the parameters with a decay factor of 0.999 to stabilize training. For short-time Fourier transform, the FFT size and hop length are set to 510 and 128, respectively. The source code is available online\footnote{\url{https://github.com/zhangwen0821/ARFSE.git}}.

Our evaluation proceeds in two stages. First, we evaluate ARFSE on VoiceBank+DEMAND to compare its core performance and inference efficiency with FlowSE and BBED under different NFE configurations. Second, to test generalization and robustness in unseen acoustic environments, we benchmark the models on the DNS Challenge dataset. 

Performance is evaluated using standard metrics including PESQ~\cite{recommendation2001perceptual} scores,  wav2vec MOS (WVMOS)~\cite{andreev2023hifi++}, eSTOI~\cite{jensen2016algorithm}, DNSMOS P.835~\cite{reddy2022dnsmos} including SIG, BAK, OVRL, Scale-Invariant Signal-to-Distortion Ratio (SI-SDR)~\cite{le2019sdr} and Real-Time Factor (RTF). The RTF is defined as:
\begin{equation}
    \text{RTF} = \frac{T_{\text{infer}}}{T_{\text{audio}}}
\end{equation}
where $T_{\text{infer}}$ denotes the total wall-clock inference time required by the model, and $T_{\text{audio}}$ represents the actual physical duration of the processed speech signals.

\begin{table*}[t]
  \caption{Experimental results on the VoiceBank+DEMAND dataset. $\uparrow$ denotes higher metric value indicates better performance. Bold and underlined values represent the best overall results and the best results within a specific method.}
  \label{tab:vb_dmd}
  \centering
  \footnotesize
  \setlength{\tabcolsep}{7pt}
  \renewcommand\arraystretch{1.1}
  \begin{tabular}{l c c c c c c c c c}
    \toprule
    \textbf{Method} & \textbf{NFEs} & \textbf{PESQ} $\uparrow$ & \textbf{eSTOI} $\uparrow$ & \textbf{SI-SDR} $\uparrow$ & \textbf{WV-MOS} $\uparrow$ & \textbf{DNSMOS} $\uparrow$ & \textbf{SIG} $\uparrow$ & \textbf{BAK} $\uparrow$ & \textbf{OVRL} $\uparrow$ \\
    \midrule
    \multirow{2}{*}{BBED~\cite{lay23_interspeech}}
    & $60$ & \underline{$3.09$} & \underline{$0.88$} & \underline{$18.75$} & \underline{$4.29$} & \underline{$3.57$} & $3.48$ & \underline{$4.04$} & \underline{$3.20$} \\
    & $1$ & $2.43$ & $0.86$ & $16.84$ & $3.93$ & $3.30$ & $3.48$ & $3.68$ & $3.02$ \\
    \midrule
    \multirow{3}{*}{FLOWSE~\cite{10888274}}
    & $5$ & \underline{$3.05$} & $0.87$ & $18.91$ & \underline{$\mathbf{4.30}$} & $3.56$ & \underline{$3.48$} & $4.04$ & $3.20$ \\
    & $2$ & $2.89$ & $0.87$ & \underline{$19.67$} & $4.29$ & \underline{$3.57$} & $3.47$ & \underline{$\mathbf{4.08}$} & \underline{$3.21$} \\
    & $1$ & $2.86$ & $0.87$ & $19.57$ & $4.26$ & $3.56$ & $3.46$ & $4.06$ & $3.19$ \\
    \midrule
    \multirow{3}{*}{ARFSE}
    & $5$ & \underline{$\mathbf{3.11}$} & $0.88$ & $18.02$ & $4.27$ & $3.56$ & \underline{$\mathbf{3.50}$} & $4.03$ & $3.20$ \\
    & $2$ & $3.06$ & $0.88$ & $19.19$ & \underline{$4.29$} & $3.57$ & $3.49$ & $4.06$ & $3.21$ \\
    & $1$ & $3.00$ & $0.88$ & \underline{$\mathbf{19.91}$} & $4.27$ & \underline{$\mathbf{3.58}$} & $3.48$ & \underline{$4.07$} & \underline{$\mathbf{3.22}$} \\
    \bottomrule
  \end{tabular}
\end{table*}

\begin{table*}[t]
\centering
\caption{The comparison of generalization capabilities of FLOWSE and ARFSE. The models are trained on VoiceBank+DEMAND and tested on DNS Challenge dataset.}
\label{tab:generalization_ablation}
\footnotesize
\setlength{\tabcolsep}{7pt}
\renewcommand\arraystretch{1.1}
\begin{tabular}{l c c c c c c c c c }
\toprule
\multirow{2}{*}{\textbf{Method}} & \multirow{2}{*}{\textbf{NFEs}} & \multicolumn{4}{c}{\textbf{no reverb}} & \multicolumn{4}{c}{\textbf{with reverb}} \\
\cmidrule(lr){3-6} \cmidrule(lr){7-10}
& \textbf{} & \textbf{PESQ} $\uparrow$ & \textbf{eSTOI} $\uparrow$ & \textbf{SI-SDR} $\uparrow$ & \textbf{DNSMOS} $\uparrow$ & \textbf{PESQ} $\uparrow$ & \textbf{eSTOI} $\uparrow$ & \textbf{SI-SDR} $\uparrow$ & \textbf{DNSMOS} $\uparrow$ \\
\midrule
\multirow{3}{*}{FLOWSE}
& $5$ & \underline{$\mathbf{2.39}$} & $0.90$ & $15.82$ & \underline{$4.03$} & \underline{$\mathbf{1.09}$} & $0.36$ & $-0.98$ & \underline{$3.48$} \\
& $2$ & $2.25$ & $0.90$ & \underline{$16.12$} & $4.01$ & $1.08$ & $0.36$ & $-0.26$ & $3.31$ \\
& $1$ & $2.25$ & $0.90$ & $16.04$ & $3.98$ & $1.08$ & \underline{$\mathbf{0.37}$} & \underline{$\mathbf{-0.21}$} & $3.26$ \\
\midrule
\multirow{3}{*}{ARFSE}
& $5$ & \underline{$2.38$} & \underline{$\mathbf{0.92}$} & $16.59$ & \underline{$\mathbf{4.04}$} & \underline{$\mathbf{1.09}$} & \underline{$0.35$} & $-1.17$  & \underline{$\mathbf{3.55}$} \\
& $2$ & $2.31$ & $0.91$ & $17.07$ & $4.02$ & $1.08$ & $0.34$ & $-1.20$ & $3.36$ \\
& $1$ & $2.26$ & $0.91$ & \underline{$\mathbf{17.07}$} & $4.00$ & $1.07$ & $0.34$ & \underline{$-0.92$} & $3.28$ \\
\bottomrule
\end{tabular}
\vspace{-8pt}
\end{table*}

\begin{table}[t]
\centering
\caption{Ablation study of explicit timestep embedding on VoiceBank+DEMAND dataset.}
\label{tab:ablation}
\footnotesize
\setlength{\tabcolsep}{3pt}
\renewcommand\arraystretch{1.1}
\begin{tabular}{lccccc}
\toprule
\textbf{Method} & \textbf{NFEs} & \textbf{PESQ} $\uparrow$ & \textbf{eSTOI} $\uparrow$ & \textbf{SI-SDR} $\uparrow$ & \textbf{RTF} $\downarrow$ \\
\midrule
\multirow{2}{*}{FlowSE (w/ $t$)}
& $5$ & $2.98$ & $0.87$ & $18.46$ & $0.19$ \\
& $1$ & $2.87$ & $0.87$ & $20.02$ & $0.05$ \\
\addlinespace
\midrule
\multirow{2}{*}{ARFSE (w/o $t$)}
& $5$ & $3.09$ & $0.88$ & $18.42$ & $0.13$ \\
& $1$ & $2.97$ & $0.88$ & $20.01$ & $0.02$ \\
\addlinespace
\midrule
MeanFlowSE~\cite{11465151} & $1$ & $2.94$ & $0.88$ & $19.97$ & $0.11$ \\
\bottomrule
\end{tabular}
\vspace{-10pt}
\end{table}

\section{Results}
\subsection{Performance on VoiceBank+DEMAND Dataset}
As shown in Table~\ref{tab:vb_dmd}, ARFSE delivers competitive performance against baseline methods across different NFE configurations. At $NFE=5$, ARFSE achieves a PESQ of 3.11 and an eSTOI of 0.88, compared to a PESQ of 3.05 for FlowSE. Under the single-step constraint ($NFE=1$), ARFSE maintains a PESQ of 3.00 and an eSTOI of 0.88, while the single-step BBED and FlowSE yield PESQ scores of 2.43 and 2.86, respectively. Furthermore, single-step ARFSE yields an SI-SDR of 19.91~dB, a DNSMOS of 3.58, and an OVRL of 3.22. These results indicate that the autonomous dynamical system mitigates the performance degradation typically observed in time-conditioned generative frameworks under low NFE budgets.

\subsection{Generalization and Robustness}
Table~\ref{tab:generalization_ablation} compares the cross-dataset generalization capabilities of the models on the unseen DNS Challenge dataset. In this out-of-distribution evaluation, ARFSE demonstrates comparable overall performance to FlowSE. Specifically, across both the non-reverberant (\textit{no reverb}) and with reverberant (\textit{with reverb}) settings, both models exhibit similar variations in perceptual scores and objective metrics, with both frameworks exhibiting noticeable performance degradation in severe reverberant environments. These results indicate that time-unconditional framework achieves generalization comparable to traditional flows. To further validate our framework, future work will focus on extensive evaluations in universal speech enhancement.

\subsection{Ablation Study on Timestep Embedding}
To evaluate the explicit impact of the timestep embedding and inference efficiency, an ablation study was conducted within a unified 27.8M parameter framework. As reported in Table~\ref{tab:ablation}, removing the timestep embedding (w/o $t$) improves both enhancement quality and inference velocity.

At $NFE=1$, ARFSE (w/o $t$) achieves a PESQ of 2.97 and an eSTOI of 0.88, whereas FlowSE (w/ $t$) and the MeanFlowSE baseline yield PESQ scores of 2.87 and 2.94, respectively. Concurrently, by removing the architectural modules associated with temporal conditioning and time-step modulation layers, ARFSE reduces the RTF to 0.02 at $NFE=1$. This corresponds to a lower RTF compared to FlowSE (0.05) and MeanFlowSE (0.11), providing empirical support for the hypothesis that explicit temporal conditioning is mathematically redundant for linear-path speech enhancement.

\section{Conclusion}
In this paper, we proposed a novel time-unconditional generative speech enhancement method based on autonomous rectified flow. By formulating the enhancement process as an autonomous ODE system, the proposed method eliminates the need for explicit time-step embeddings. We showed that, under a linear interpolation path, the target vector field is inherently time-invariant and equivalent to the underlying noise distribution. Experimental results on VoiceBank+DEMAND dataset demonstrate that ARFSE achieves competitive speech quality with $NFE=5$ (PESQ 3.11) and delivers top-tier performance for single-step generation ($NFE=1$), with a remarkably low RTF of 0.02. In future work, we will further investigate the effectiveness of ARFSE in universal speech enhancement tasks.

\newpage

\section{Acknowledgments}
This work was supported in part by Yangtze River Delta Science and Technology Innovation Community Joint Research under Grant 2024CSJGG1100 and in part by the Zhejiang Provincial Natural Science Foundation of China (No. LMS26F020008).

\section{Generative AI Use Disclosure}
We disclose that generative AI tools were used solely for language polishing, text refinement, and format checking of this manuscript, and were not used to generate any core research content or substantive academic results. 

\bibliographystyle{IEEEtran}
\bibliography{refs}
\end{document}